\newcommand{\lmax}{\ell_{\mathrm{max}}}
\newcommand{\Nbs}{N_{\scriptstyle\mathrm{bs}}}
\newcommand{\FigureLength}{6.2cm}

\documentclass[floatfix,aps,prb,twocolumn,showpacs,notitlepage,superscriptaddress]{revtex4-1}
\usepackage{graphicx,amsmath,mathrsfs}
\usepackage{amssymb}
\usepackage{amsthm}
\usepackage{bm}
 
\begin{document}
\title{First-principles density-functional calculations using
localized spherical-wave basis sets}
\author{C.~K.~Gan, P.~D.~Haynes, and M.~C.~Payne}
\address{Theory of Condensed Matter, Cavendish Laboratory, Madingley
Road, Cambridge CB3~0HE, United Kingdom}
 
\begin{abstract}
We present a detailed study of the use of localized spherical-wave
basis sets, first introduced in the context of linear-scaling, in
first-principles density-functional calculations.  Several parameters
that control the completeness of this basis set are fully
investigated on systems such as molecules and bulk crystalline
silicon.  We find that the results are in good agreement with those
obtained using the extended plane-wave basis set.  Since the
spherical-wave basis set is accurate, easy to handle, relatively
small, and can be systematically improved, we expect it to be of use
in other applications.
\end{abstract}
\maketitle
 
\section{Introduction}
\par Localized basis sets such as Gaussians\cite{Boys_50v200}, 
truncated pseudo-atomic
orbitals\cite{Sankey_89v40,Artacho_99v215}, real-space
grids\cite{Chelikowsky_94v72,Chelikowsky_94v50,Modine_97v55,Bernholc_97v65,Fattebert_00v62,Beck_00v72}, B-spline (or ``blip'')
functions\cite{Hernandez_97v55}, and
wavelets\cite{Arias_99v71}, have been used in first-principles
calculations. Recently there has been a surge of activity to
investigate linear-scaling methods\cite{Goedecker_99v71} (where the
computational effort and memory requirement scale linearly with the
system size), all of which use localized basis sets in their
implementations.  One localized basis set that was introduced for
linear-scaling methods, the spherical-wave basis set\cite{Haynes_97v102},
is interesting because while sharing some of the properties (such as
the concept of energy cutoff) with the extended plane-wave basis set,
it possesses other advantages such as each basis function being fully
localized within a sphere.  Even though it has been used to implement
a linear-scaling method and tested against bulk crystalline
silicon\cite{Haynes_99v59}, this basis set has not yet been fully
investigated.  This work serves to reveal the properties of this
localized basis set using a matrix diagonalization approach, which
frees us from having to consider other sources of error introduced by
other cutoffs (such as the density-matrix spatial
cutoff\cite{Li_93v47,Hernandez_96v53,Haynes_98v108}).  The completeness and
appropriateness of this basis set are investigated in first-principles
calculations within density-functional theory through applications to
molecules and bulk crystalline silicon.  The remainder of this work is
organized as follows. Section~\ref{sec_origin} introduces the
spherical-wave basis set.  In section~\ref{sec_dfttheory}, a brief
description is given of the first-principles calculation within
density-functional theory using the spherical-wave basis set.
Section~\ref{sec_results} contains the results of calculations on
different test systems, which are compared with those obtained using
the same theory level and approximations, but with a plane-wave basis
set\cite{Payne_92v64}.  Section~\ref{sec_summary} contains the summary
and conclusions.

\section{Origin of the basis set}
\label{sec_origin}
\par The spherical-wave basis functions\cite{Haynes_97v102} used in this
work are eigenfunctions of the Helmholtz equation
\begin{equation}
(\nabla^2 + q^2) \chi({\bf r}) = 0,
\label{eq_Helmholtz}
\end{equation}
subject to boundary conditions such that the solutions $\chi({\bf r})$ are
nonvanishing only inside a sphere of radius $R$ and vanishing
whenever $ |{\bf r}| \ge R $.  The eigenfunctions are
\begin{equation}
\chi(r,\theta,\phi) = \left\{
\begin{array}{ll}
j_{\ell}(q_{n \ell} r) Y_{\ell m}(\theta,\phi), & r < R,
\nonumber\\
0, & r \ge R,
\end{array}
\right.
\label{eq_eigen}
\end{equation}
where $(r,\theta,\phi)$ are spherical polar coordinates with origin at
the center of the sphere, $\ell$ is a non-negative integer and $m$ is
an integer satisfying $-\ell \le m \le \ell$.  $j_{\ell}(x)$ is the
spherical Bessel function of order $\ell$, and $Y_{\ell m}(\theta,\phi)$
is a spherical harmonic.  The eigenvalue $q_{n \ell}$ is determined from
the $n$th zero of $j_{\ell}(x)$ where
\begin{equation}
j_{\ell}(q_{n \ell} R) = 0.
\end{equation}
We note that each eigenfunction in Eq.~(\ref{eq_eigen}) has an energy
of $\hbar^2 q_{n \ell}^2/(2m_{\mathrm{e}})$, hence it is possible to
use the concept of cutoff energy to restrict the number of $q_{n
\ell}$ in the expansion of a wavefunction.

\par The real spherical-wave basis functions used in this work are
\begin{equation}
\chi_{\alpha,n \ell m}({\bf r}) = \left\{
\begin{array}{ll}
j_{\ell}(q_{\alpha,n \ell}
|{\bf r}-{\bf R}_{\alpha}|)\overline{Y}_{\ell m}
(\Omega_{{\bf r}-{\bf R}_{\alpha}}),
& |{\bf r}-{\bf R}_{\alpha}| < r_{\alpha},
\nonumber\\
0, & |{\bf r}-{\bf R}_{\alpha}| \ge r_{\alpha} ,
\end{array}
\right.
\label{eq_neweigen}
\end{equation}
where $\alpha$ signifies a {\em basis sphere} with radius $r_{\alpha}$
and centered at ${\bf R}_{\alpha}$. $\overline{Y}_{\ell m}
(\theta,\phi)$ are the real linear combinations of the spherical
harmonics.  By construction, all basis functions within a basis sphere
are orthogonal to one another.  In general, more than one basis sphere
is needed to expand a wavefunction
\begin{equation}
\psi({\bf r}) =
\sum_{\alpha, n \ell m} c_{\alpha, n \ell m}
\chi_{\alpha, n \ell m}({\bf r}),
\label{eq_basis_expansion}
\end{equation}
where $c_{\alpha, n \ell m}$ are the associated coefficients. 
For most systems tested in this work, we have 
used one basis sphere per atom,
where the basis spheres are centered
on the atoms. For some systems we have increased the number of
basis spheres by placing basis spheres between the atoms. 
In principle it is possible to use two or more basis spheres of
different radii centered
on the same atom, but this arrangement has not been studied in this work.
We note that even
though the basis functions belonging to different basis spheres are
generally nonorthogonal, one of the main advantages of this basis set
is that it is possible to analytically evaluate\cite{Haynes_97v102} the
overlap matrix elements
\begin{equation}
S_{\alpha, n \ell m; \beta, n' \ell' m'} =
\int d{\bf r}\  \chi_{\alpha, n \ell m}({\bf r}) \chi_{\beta, n' \ell'
m'}({\bf r}),
\label{eq_S}
\end{equation}
and kinetic energy matrix elements
\begin{equation}
T_{\alpha, n \ell m; \beta, n' \ell' m'} = -\frac{\hbar^2}{2
m_{\mathrm{e}}} \int d{\bf r}\ \chi_{\alpha, n \ell m}({\bf r}) \nabla^2
\chi_{\beta, n' \ell' m'}({\bf r}).
\label{eq_T}
\end{equation}
We also note that the matrix elements for the nonlocal
pseudopotentials in the Kleinman-Bylander\cite{Kleinman_82v48} form can
also be evaluated analytically by first expanding the projectors in
the spherical-wave basis set.
 
\section{Density-functional calculations}
\label{sec_dfttheory}
\par The Kohn-Sham (KS) equation for an $M$-electron system
is\cite{Hohenberg_64v136,Kohn_65v140}
\begin{equation}
\hat{H} \psi_m({\bf r}) = \left[-\frac{\hbar^2}{2m_{\mathrm{e}}}\nabla^2
+ V_{\mathrm{eff}}({\bf r}) \right] \psi_m({\bf r}) =
\varepsilon_m \psi_m({\bf r}),
\label{eq_KS}
\end{equation}
where $\{ \psi_m({\bf r}) \}$ are the KS eigenfunctions with
corresponding eigenvalues $\{ \varepsilon_m \}$. The effective potential
$V_{\mathrm{eff}}({\bf r})$ consists of the classical electrostatic
potential, the ionic potential due to the nuclei and the
exchange-correlation potential.  The effective potential depends on
the electron density, $\rho({\bf r})$, which is formed from the $M$
lowest eigenstates
\begin{equation}
\rho({\bf r}) = \sum_{m=1}^{M} \left| \psi_m({\bf r}) \right|^2.
\end{equation}
 
\par We use the real spherical-wave basis set $\{\chi_{\nu}({\bf
r})\}$ to expand the $n$-th KS eigenstate
\begin{equation}
\psi_n({\bf r}) = \sum_{\nu} {x_n}^{\nu} \ \chi_{\nu}({\bf r}),
\label{eq_expand}
\end{equation}
where $\nu$ is a collective label for $(\alpha, n \ell m)$ in
Eq.~(\ref{eq_basis_expansion}).  Substituting Eq.~(\ref{eq_expand})
into Eq.~(\ref{eq_KS}), and taking inner products with the $\{
\chi_{\mu}({\bf r}) \}$, we obtain the generalized eigenvalue problem
\begin{equation}
\sum_{\nu} (H_{\mu \nu} - \varepsilon_n S_{\mu \nu}) 
{x_n}^{\nu} = 0,
\label{eq_HxeSx}
\end{equation}
where the overlap matrix is given by
\begin{equation}
S_{\mu \nu} = \int d{\bf r}\ \chi_{\mu}({\bf r})
\chi_{\nu}({\bf r}),
\end{equation}
and the Hamiltonian matrix is given by
\begin{equation}
H_{\mu\nu} = \int d{\bf r}\ \chi_{\mu}({\bf r}) \hat{H}
\chi_{\nu}({\bf r}).
\end{equation}
 
\par It should be emphasized that when a large system is
studied, $S$ and $H$ will be sparse.  In this case it is more
efficient to use an iterative method based on preconditioned
conjugate gradient minimization\cite{Gan_01v134} to find the lowest few
eigenvalues and corresponding eigenvectors of Eq.~(\ref{eq_HxeSx})
than to use a direct matrix diagonalization
method~\cite{Golub_96,Press_86} in which all eigenvalue-eigenvector
pairs are found.
 
\par The completeness of the basis set depends on several parameters
such as the radius of the basis sphere, $R$; the maximum angular
momentum component, $\lmax$; and the number of $q_{n \ell}$ values for
each angular momentum component, which we will take here to be the
same for all $\ell$ and is denoted by $n_q$.  The number of basis
functions in a basis sphere is $(\lmax+1)^2 n_q$. For a fixed number
of $n_q$, we note that the number of basis functions increases very
rapidly with respect to $\lmax$.  However, we will demonstrate that
most physical properties can be deduced using only a small $\lmax$,
which is typically 2. The cutoff energy $E_{\mathrm{c}}$ for a basis
sphere is roughly given by
\begin{equation}
E_{\mathrm{c}} = \frac{\hbar^2}{2m_{\mathrm{e}}} \left(
\frac{n_q \pi}{R} \right)^2.
\end{equation}
Periodicity of the system under study is assumed and the $\Gamma$
point only is used for the Brillouin zone sampling.  We have used the
local density approximation (LDA) for the exchange and correlation
term. Norm-conserving Troullier-Martins\cite{Troullier_91v43}
pseudopotentials in the Kleinman-Bylander\cite{Kleinman_82v48} form are
used.
 
\section{Results of the calculations}
\label{sec_results}
\par In this section we present and discuss the results obtained from
density-functional calculations using the spherical-wave basis set. We
study the convergence of the total energy as a function of the cutoff
energy, $E_{\mathrm{c}}$; the radii of the basis spheres, $R$; the
maximum angular momentum component, $\lmax$; and the number of basis
spheres, $\Nbs$. Physical properties are deduced from total energy
calculations on the systems. For molecules, we calculate the
equilibrium bond lengths and force constants. For bulk crystalline
silicon, we calculate the equilibrium lattice parameter and bulk
modulus. These results are compared with those obtained using a
plane-wave code\cite{Payne_92v64}, and from experiment\cite{CRC_95}.
 
%----------
\begin{figure}[htbp]
%\centerline{\includegraphics[width=8cm,clip]{./fig.01.eps}}
\includegraphics[width=\FigureLength,clip]{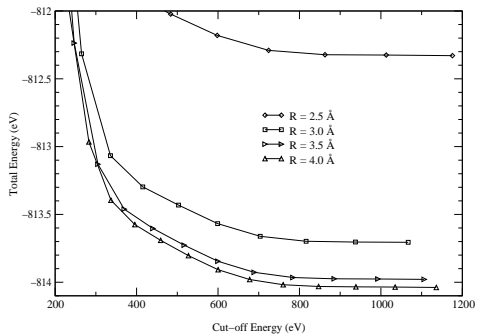}
\caption{Total energy of the chlorine molecule with a bond length of
1.6~\AA.  Two basis spheres of radius $R$ centered on the atoms are
used. The cubic simulation cell has sides of length 12~\AA.  $\lmax = 2$.
}
\label{fig_box12cl2BL1_6}
\end{figure}
%==================
\par In Fig.~\ref{fig_box12cl2BL1_6} we plot the total energy of the
chlorine molecule with a bond length of 1.6~\AA\ as a function of
cutoff energy $E_{\mathrm{c}}$ and basis sphere radius $R$.  The
figure shows that the total energy decreases rapidly as the cutoff
energy and the basis sphere radius are increased, which is to be
expected from the additional variational freedom that is introduced.
Convergence in the total energy is achieved for cutoff energies above
800~eV.
 
%----------
\begin{figure}[htbp]
%\centerline{\includegraphics[width=8cm,clip]{./fig.02.eps}}
\includegraphics[width=\FigureLength,clip]{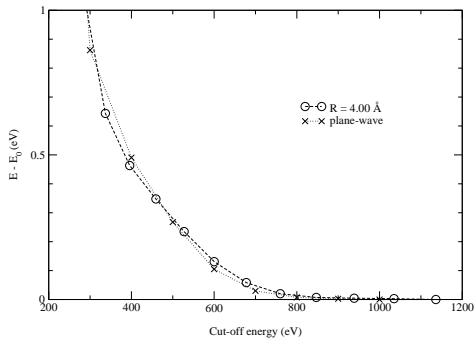}
\caption{ Nature of the convergence of the total energy of the chlorine molecule
with a bond length of 1.6~\AA. The data for the spherical-wave basis
set are taken from Fig.~\ref{fig_box12cl2BL1_6}. 
The respective converged total energies
$E_0$ are subtracted from the total energy $E$ in each case.
}
\label{fig_etot_ecutconver}
\end{figure}
%==================
\par Fig.~\ref{fig_etot_ecutconver} shows that the rate of convergence
of the total energy with respect to the cutoff energy is the same for
both the localized spherical-wave and extended plane-wave basis sets.
This confirms that the energy cutoff concept can be equally applied
in the spherical-wave basis set.
 
%----------
\begin{figure}
%\centerline{\includegraphics[width=8cm,clip]{./fig.03.eps}}
\includegraphics[width=\FigureLength,clip]{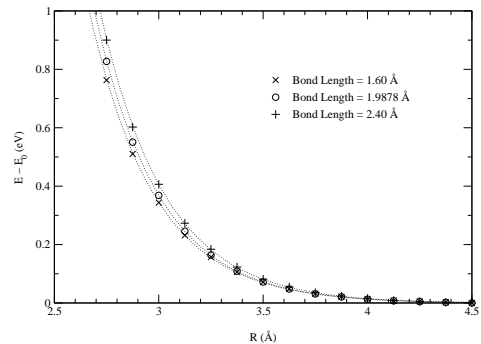}
\caption{Convergence of the total energy of the chlorine molecule for
a variety of bond lengths.  The respective converged total energies
$E_0$ are subtracted from the total energy $E$ in each case.  Two
basis spheres of radius $R$ centered on the atoms are used.  The
dotted curves are exponential fits to the data.  }
\label{fig_towards_0}
\end{figure}
%===================
 
\par Using an energy cutoff above 900~eV, we calculate the total
energy of the chlorine molecule for a variety of bond lengths as a
function of the basis sphere radius $R$.  Fig.~\ref{fig_towards_0}
shows that the total energy converges exponentially with respect to
$R$.  We also note that the total energy converges slightly faster
with respect to $R$ for molecules with smaller bond lengths. This
reflects the fact that for a given $R$, the basis set is more complete
for a smaller molecule than a larger one because the basis spheres are
closer to one another in the smaller molecule.

%----------
\begin{figure}[htbp]
%\centerline{\includegraphics[width=8cm,clip]{./fig.04.eps}}
\includegraphics[width=\FigureLength,clip]{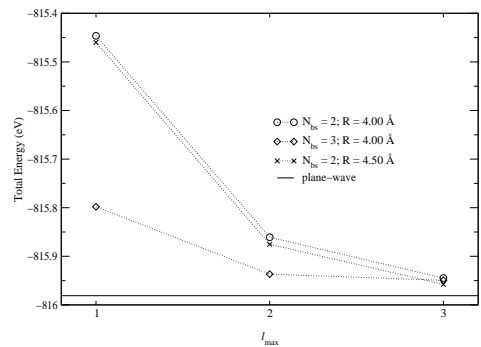}
\caption{
Total energy of the chlorine molecule with a bond length
of $2.4$~\AA\ as a function of the maximum angular momentum component
$\lmax$ for different basis-sphere radii $R$ and 
numbers of basis spheres $\Nbs$.
For the three-basis-sphere calculations, two basis spheres
are centered on the atoms and a third basis sphere is centered between
the atoms. 
For each spherical-wave calculation, we have used a value of $n_q$ which
is the smallest integer such that the cutoff energy exceeds 900~eV.
The horizontal solid line
corresponds to the 
total energy obtained from the plane-wave calculation with a cutoff
energy of 900~eV.
}
\label{etot_lmax}
\end{figure}
%===================
\par 
Since the total energy also depends on other parameters such as
$l_{\rm{max}}$ and the number of basis spheres $N_{\rm{bs}}$, we
have
performed calculations on the chlorine molecule with a bond length
of
2.4~\AA. The results in Fig.~\ref{etot_lmax} show the convergence of the total
energy of the system as a function of $\lmax$ for different
basis-sphere radii $R$ and numbers of basis spheres $\Nbs$.
The
rapid convergence of the total energy with respect to
$\lmax$
is evident from the figure. We note that the ``best'' result
obtained
from the spherical-wave calculation with $\Nbs=2$, 
$R=$~4.50~\AA, and $\lmax=3$ gives a total energy of
$-815.958$~eV, which lies 0.023~eV above the plane-wave total
energy
of $-815.981$~eV. This difference, which is due to the
incompleteness
of the spherical-wave basis set, could be reduced further by
increasing the basis-sphere radius $R$ and $\lmax$.
However, we are content with this accuracy because the error due to the
incompleteness of the spherical-wave basis set is only about
$3\times 10^{-5}$ of the total energy obtained from the plane-wave
calculation.
The number of spherical-wave basis functions in this case is only
672, which is a small fraction (0.6\%) of 112452, the number of plane
waves.

%--------------
\begin{figure}[t]
%\centerline{\includegraphics[width=8cm,clip]{./fig.05.eps}}
\includegraphics[width=\FigureLength,clip]{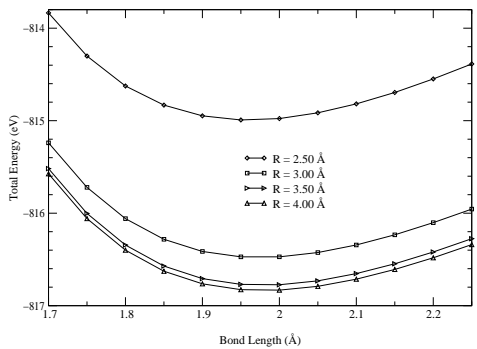}
\caption{Total energy of the chlorine molecule as a function of the
bond length.  Two basis spheres of the same radius $R$ centered on the
atoms are used.  $\lmax = 2$.}
\label{fig_cl2_800eV}
\end{figure}
%=================

%-------------
\begin{table}[h]
\begin{tabular}{c|rrrr|rrrr}
& \multicolumn{4}{c|}{$\Nbs$ = 2} & \multicolumn{4}{c}{$\Nbs$ = 3} \\
$R$(\AA)   & $r_{\mathrm{e}}$(\AA)  & $\delta r_{\mathrm{e}}(\%) $ &
$f$$\left(\frac{\mathrm{N}}{\mathrm{cm}}\right)$
& $\delta f(\%)$ & $r_{\mathrm{e}}$(\AA) &  $\delta r_{\mathrm{e}}(\%) $ &
$f$$\left(\frac{\mathrm{N}}{\mathrm{cm}}\right)$ 
& $\delta f(\%) $\\
\hline
 2.00    & 1.8941 & $-$3.66  & 5.276 & 39.2 & 
 1.9158 & $-$2.56 & 4.591 & 21.1 \\
 2.50    & 1.9609 & $-$0.26  & 3.959 & 4.5 & 
 1.9537 & $-$0.63 & 4.017 & 6.0\\
 3.00    & 1.9733 & 0.37     & 3.754 & $-$0.9 &
 1.9636 & $-$0.13 & 3.849 & 1.6\\
 3.50    & 1.9769 & 0.55     & 3.694 & $-$2.5  & 
 1.9658 & $-$0.02 & 3.802 & 0.3\\
 4.00    & 1.9777 & 0.59     & 3.673 & $-$3.1  & 
 1.9663 & 0.01 & 3.789 & 0.0\\
 4.50    & 1.9778 & 0.60     & 3.670 & $-$3.2  & 
 1.9663 & 0.01 & 3.786  & $-$0.1\\
\end{tabular}
\caption{ Results for the equilibrium bond length $r_{\mathrm{e}}$ and
force constant $f$ of the chlorine molecule with $\lmax$ = 2.  The
meanings of $R$ and $\Nbs$ are explained in the caption of Table
\ref{tb_cl2_800eV_lmax1}.  }
\label{tb_cl2_800eV_lmax2}
\end{table}
%==================
\par To study the effect of $\Nbs$, $\lmax$ and $R$ on the calculated
physical properties such as the equilibrium bond length
$r_{\mathrm{e}}$ and force constant $f$, we perform a series of
calculations on the chlorine molecule for a range of bond lengths from
1.70~\AA\ to 2.25~\AA.  A typical result is shown in
Fig.~\ref{fig_cl2_800eV}.  The results of the calculations of
$r_{\mathrm{e}}$ and $f$ with $\lmax = $ 1, 2, and 3 are displayed in
Tables~\ref{tb_cl2_800eV_lmax1}, \ref{tb_cl2_800eV_lmax2}, and
\ref{tb_cl2_800eV_lmax3}, respectively.  The errors in
$r_{\mathrm{e}}$ and $f$ displayed in the columns headed under $\delta
r_{\mathrm{e}}$ and $\delta f$ are deduced from the results of the
plane-wave calculations.

%-------------
\begin{table}[p]
\begin{tabular}{c|lrrr|rrrr}
& \multicolumn{4}{c|}{$\Nbs$ = 2}& \multicolumn{4}{c}{$\Nbs$ = 3} \\
$R$(\AA) & $r_{\mathrm{e}}$(\AA) & $\delta r_{\mathrm{e}}(\%) $ &
$f$$\left(\frac{\mathrm{N}}{\mathrm{cm}}\right)$ & $\delta f(\%)$ &
$r_{\mathrm{e}}$(\AA) & $\delta r_{\mathrm{e}}(\%) $ &
$f$$\left(\frac{\mathrm{N}}{\mathrm{cm}}\right)$ & $\delta f(\%) $\\
\hline 2.00 & 2.0214 & 2.81& 4.880 & $ $28.8 & 1.9565& $-$0.49 & 4.492
& $ $18.5\\ 2.50 & 2.1240 & 8.03& 2.542 & $-$32.9 & 1.9988& $ $1.66 &
3.747 & $-$1.1\\ 3.00 & 2.1536 & 9.54& 1.874 & $-$50.6 & 2.0098& $
$2.22 & 3.557 & $-$6.1\\ 3.50 & 2.1595 & 9.84& 1.772 & $-$53.2 &
2.0123& $ $2.35 & 3.496 & $-$7.8\\ 4.00 & 2.1623 & 9.98& 1.703 &
$-$55.1 & 2.0127& $ $2.37 & 3.484 & $-$8.1\\ 4.50 & 2.1625 & 9.99&
1.701 & $-$55.1 & 2.0128& $ $2.38 & 3.481 & $-$8.2\\
\end{tabular}
\caption{ Results for the equilibrium bond length $r_{\mathrm{e}}$
and force constant $f$ of the chlorine molecule with $\lmax$ = 1.
When $\Nbs=2$, two basis spheres of the same radius $R$ centered on the
atoms are used. When $\Nbs=3$, three basis spheres of the same radius $R$
are used, the third basis sphere being centered between the atoms.
The experimental values for $r_{\mathrm{e}}$ and $f$ are 1.9878~\AA\
and 3.23~N/cm, respectively. The plane-wave calculations give values
of 1.9661~\AA\ and 3.790~N/cm, respectively, where we have used the
same pseudopotentials, Brillouin zone sampling, and cutoff energy.  }
\label{tb_cl2_800eV_lmax1}
\end{table}
%==================
\par In Table~\ref{tb_cl2_800eV_lmax1}, the values of $r_{\mathrm{e}}$
and $f$ converge rapidly with respect to $R$.  However, the results
with two basis spheres and $\lmax=1$ shows that the converged results
contain unacceptably large systematic errors.  The inclusion of a
third sphere reduces the errors significantly because the bonding
region between the atoms is described better by the third sphere.  The
results show it is impossible to improve the results simply by
enlarging $R$ when $\lmax=1$ is used.

\par We repeat the calculations for $r_{\mathrm{e}}$ and $f$ with
$\lmax=2$, for which the results are presented in
Table~\ref{tb_cl2_800eV_lmax2}.  The converged results with $\Nbs=2$ and
$\lmax=2$ are better than the converged results with $\Nbs=3$ and
$\lmax=1$, which indicates the importance of $\lmax$ over $\Nbs$ for the
``minimal basis set'' calculations. With $\lmax=2$ and $\Nbs=2$, the error
of the converged results for $r_{\mathrm{e}}$ and $f$ are $-$0.50\%
and 13.6\% compared to the experimental values, respectively.  These
accuracies are acceptable within the LDA.

%-------------
\begin{table}[p]
\begin{tabular}{l|rrrr}
$R$(\AA)   & $r_{\mathrm{e}}$(\AA)  & $\delta r_{\mathrm{e}}(\%) $ &
$f$$\left(\frac{\mathrm{N}}{\mathrm{cm}}\right)$
& $\delta f(\%)$ \\
\hline
 2.00    & 1.8833 & $-$4.21  & 5.212 &  $ $37.5  \\
 2.50    & 1.9481 & $-$0.92  & 4.128 &  $ $8.9  \\
 3.00    & 1.9636 & $-$0.13  & 3.860 &  $ $1.8 \\
 3.50    & 1.9668 & $ $0.04  & 3.792 &  $ $0.1 \\
 4.00    & 1.9674 & $ $0.07  & 3.777 &  $-$0.3 \\
 4.50    & 1.9675 & $ $0.07  & 3.773 &  $-$0.4 \\
\end{tabular}
\caption{ Results for the equilibrium bond length $r_{\mathrm{e}}$
and force constant $f$ of the chlorine molecule with $\lmax$ = 3.  }
\label{tb_cl2_800eV_lmax3}
\end{table}
%==================
\par Finally in Table~\ref{tb_cl2_800eV_lmax3}, we present the values
of $r_{\mathrm{e}}$ and $f$ using two basis spheres centered on the
atoms $\lmax=3$.  As expected, the converged values of
$r_{\mathrm{e}}$ and $f$ agree very well with the plane-wave results.
We note that calculations with $\lmax=3$ are expensive, since the
number of basis functions is almost double that for $\lmax=2$.

%---------
\begin{figure}
%\centerline{\includegraphics[width=8cm,clip]{./fig.06.eps}}
\includegraphics[width=\FigureLength,clip]{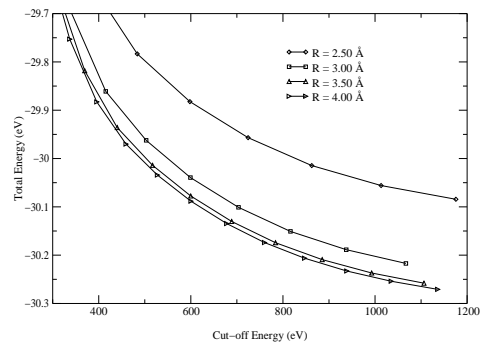}
\caption{Total energy of the hydrogen molecule with a bond length of
1.0~\AA.  Two basis spheres of the same radius $R$ centered on the
atoms are used.  $\lmax = 2$. The cubic simulation cell has sides of
length 12~\AA.  }
\label{fig_h2BL1_0}
\end{figure}
%============

\par Next we calculate the total energy of hydrogen molecule with a
bond length of 1.0~\AA\ as a function of the cutoff energy
$E_{\mathrm{c}}$, and the basis sphere radius $R$, for which the
results are displayed in Fig.~\ref{fig_h2BL1_0}.  The total energy
converges rather slowly with respect to the cutoff energy because a
bare Coulomb potential due to the hydrogen atom is used. Such behavior
is also observed in the plane-wave calculations.  However, the
convergence of energy differences is achieved when the cutoff energy
exceeds 800~eV.

%-------------
\begin{table}
\begin{tabular}{l|rrrr|rrrr}
& \multicolumn{4}{c|}{$\Nbs$ = 2} & \multicolumn{4}{c}{$\Nbs$ = 3} \\
$R$(\AA)   & $r_{\mathrm{e}}$(\AA)  & $\delta r_{\mathrm{e}}(\%) $ &
$f$$\left(\frac{\mathrm{N}}{\mathrm{cm}}\right)$
& $\delta f(\%)$ & $r_{\mathrm{e}}$(\AA) &  $\delta r_{\mathrm{e}}(\%) $ &
$f$$\left(\frac{\mathrm{N}}{\mathrm{cm}}\right)$
& $\delta f(\%) $\\
\hline
  2.00    & 0.7476    & $-$3.05 & 5.998  & 15.4 & 0.7503
  &$-$2.70& 5.865 & 12.9\\
  2.50    & 0.7643    & $-$0.88  & 5.420 &4.3  & 0.7668 &
  $-$0.56 & 5.232 & 0.7\\
  3.00    & 0.7695  & $-$0.21 & 5.237 &0.8  & 0.7712 & 0.01 &
  5.198 & 0.0\\
  3.50    & 0.7709 & $-$0.03  & 5.193 & $-$0.1  \\
  4.00    & 0.7710  & $-$0.01  & 5.178 & $-$0.4\\
\end{tabular}
\caption{ Results for the equilibrium bond length $r_{\mathrm{e}}$
and force constant $f$ of the hydrogen molecule with $\lmax$ = 2.
The meanings of $R$ and $\Nbs$ are explained in the caption of Table
\ref{tb_cl2_800eV_lmax1}. Cut-off energies above 1000~eV are used.  The
experimental values for $r_{\mathrm{e}}$ and $f$ are 0.7414~\AA\ and
5.75~N/cm respectively. The equivalent plane-wave calculations give
values of 0.7711~\AA\ and 5.197~N/cm, respectively.  }
\label{tb_h2_1000eV}
\end{table}
%======================
\par We perform a series of total energy calculations on the hydrogen
molecule for a range of bond lengths to determine the values of
$r_{\mathrm{e}}$ and $f$.  The results are tabulated in
Table~\ref{tb_h2_1000eV} and show that we can use a value of $R$ as
small as 3.00~\AA\ to obtain an accuracy of less than 1\% in
$r_{\mathrm{e}}$ and $f$ with only two basis spheres.  This should be
contrasted with the case of the chlorine molecule where with $\Nbs=2$,
$\lmax=2$, and $R = 3.00$~\AA, the values of $r_{\mathrm{e}}$ and $f$
agree only fortuitously with the plane-wave results.

%---------
\begin{figure}
%\centerline{\includegraphics[width=8cm,clip]{./fig.07.eps}}
\includegraphics[width=\FigureLength,clip]{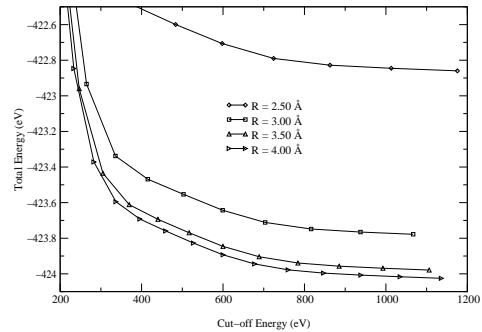}
\caption{ Total energy of the hydrogen chloride molecule with a bond
length of 1.6~\AA\ as a function of the cutoff energy and the basis
sphere radius $R$.  Two basis spheres of the same radius $R$ centered
on the atoms are used.  The cubic simulation cell has sides of length
12~\AA. $\lmax = 2$.  }
\label{fig_hcl_BL_1_60}
\end{figure}
%============
\par We can explain why, to obtain the same accuracy, the chlorine
molecule requires a larger $R$ than the hydrogen molecule.  The
equilibrium bond length of the hydrogen molecule (which is about
0.74~\AA) is smaller than the equilibrium bond length of the chlorine
molecule (which is about 1.99~\AA). The bonding region between the
hydrogen atoms is thus described better by the basis functions because
the basis spheres are closer to one another. The hydrogen molecule is
also ``smaller'' (in the sense of the extent of the charge distribution)
than the chlorine molecule.

%-------------
\begin{table}
\begin{tabular}{l|rrrr|rrrr}
& \multicolumn{4}{c|}{$\Nbs$ = 2} & \multicolumn{4}{c}{$\Nbs$ = 3} \\
$R$(\AA)   & $r_{\mathrm{e}}$(\AA)  & $\delta r_{\mathrm{e}}(\%) $ &
$f$$\left(\frac{\mathrm{N}}{\mathrm{cm}}\right)$
& $\delta f(\%)$ & $r_{\mathrm{e}}$(\AA) &  $\delta r_{\mathrm{e}}(\%) $ &
$f$$\left(\frac{\mathrm{N}}{\mathrm{cm}}\right)$
& $\delta f(\%) $\\
\hline
 2.00  & 1.2601  &$-$2.68& 6.351 &16.4& 1.2712   &$-$1.82&
 6.030  &10.5 \\
 2.50  & 1.2862  &$-$0.66& 5.683 & 4.1& 1.2885   &$-$0.49&
 5.605  &2.7 \\
 3.00  & 1.2926  &$-$0.17& 5.518 & 1.1& 1.2933   &$-$0.12&
 5.489  &0.6 \\
 3.50  & 1.2936  &$-$0.09& 5.475 & 0.3& 1.2941   &$-$0.05&
 5.461  &0.1 \\
 4.00  & 1.2942  &$-$0.05& 5.466 & 0.1& 1.2946   &$-$0.02&
 5.452  &$-$0.1 \\
 4.50  & 1.2945  &$-$0.02& 5.464 & 0.1& 1.2948   &   0.00&
 5.450  &$-$0.1 \\
\end{tabular}
\caption{Results for the equilibrium bond length $r_{\mathrm{e}}$ and
force constant $f$ of the hydrogen chloride molecule with $\lmax$ =
2.  The meanings of $R$ and $\Nbs$ are explained in the caption of Table
\ref{tb_cl2_800eV_lmax1}.  For the hydrogen chloride molecule, the
experimental values for $r_{\mathrm{e}}$ and $f$ are 1.2746~\AA\ and
5.16~N/cm, respectively. The equivalent plane-wave calculations give
values of 1.2948~\AA\ and 5.458~N/cm.  }
\label{tb_hcl_800eV}
\end{table}
%==================
\par In Fig.~\ref{fig_hcl_BL_1_60} we show the total energy of the
hydrogen chloride molecule with a bond length of 1.60~\AA\ as a
function of the cutoff energy and the radius of the basis sphere.
The energy differences converge when the cutoff energy exceeds
800~eV.  Calculations are performed to obtain $r_{\mathrm{e}}$ and
$f$, and the results are tabulated in Table~\ref{tb_hcl_800eV}.

%-------------
\begin{table}
\begin{tabular}{l|rrrr}
$R$(\AA)   & $r_{\mathrm{e}}$(\AA)  & $\delta r_{\mathrm{e}}(\%) $ &
$f$$\left(\frac{\mathrm{N}}{\mathrm{cm}}\right)$
& $\delta f(\%)$ \\
\hline
 2.00  & 1.2928 &$-$0.15& 5.508 & 0.9\\
 2.50  & 1.2937 &$-$0.08& 5.488 & 0.5\\
 3.00  & 1.2943 &$-$0.04& 5.472 & 0.3\\
 3.50  & 1.2939 &$-$0.07& 5.468 & 0.2\\
 4.00  & 1.2942 &$-$0.05& 5.466 & 0.1\\
\end{tabular}
\caption{Results for $r_{\mathrm{e}}$ and $f$ of the hydrogen chloride
molecule. Two basis spheres are used. The radius of the basis sphere
centered on the chlorine atom is fixed at 4.00~\AA\ but the radius
$R$ of the basis sphere centered on the hydrogen atom is varied.  }
\label{tb_2_diff_sp_hcl_800eV_varyrH}
\end{table}
%==================
\par We repeat the $r_{\mathrm{e}}$ and $f$ calculations for the
hydrogen chloride molecule, where the radius of the basis sphere
centered on the chlorine atom is fixed at 4.00~\AA\ but the radius of
the basis sphere centered on the hydrogen atom is varied.  The results
are presented in Table~\ref{tb_2_diff_sp_hcl_800eV_varyrH}, which
shows that we can use a smaller basis sphere of a radius of 2.0~\AA\
centered on the hydrogen atom to obtain an accuracy of less than 1\%.
It is thus possible to use different basis spheres depending on the
atomic species, which is important because this can reduce the
computation time significantly.

%----------
\begin{figure}[htbp]
%\centerline{\includegraphics[width=8cm,clip]{./fig.08.eps}}
\includegraphics[width=\FigureLength,clip]{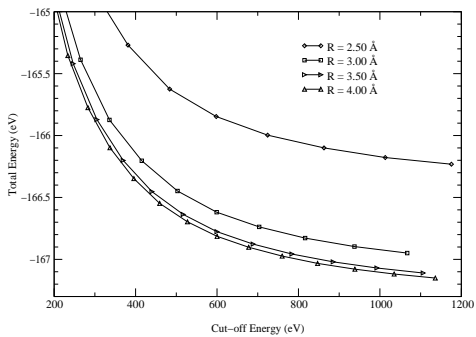}
\caption{ Total energy of the silane molecule with the Si-H bond
length of 1.83~\AA.  Five basis spheres of the same radius $R$
centered on the atoms are used.  The cubic simulation cell has a side
length of 12~\AA. $\lmax = 2$.  }
\label{fig_sih4BL1_83}
\end{figure}
%==================

%-------------
\begin{table}
\begin{tabular}{l|rrrr}
$R$(\AA)   & $r_{\mathrm{e}}$(\AA)  & $\delta r_{\mathrm{e}}(\%) $ &
$f$$\left(\frac{\mathrm{N}}{\mathrm{cm}}\right)$ & $\delta f(\%) $
\\
\hline
 2.00  & 1.4402  &$-$3.41& 14.640 & 28.6\\
 2.50  & 1.4811  &$-$0.66& 12.010 & 5.5\\
 3.00  & 1.4893  &$-$0.11& 11.523 & 1.3\\
 3.50  & 1.4906  &$-$0.03& 11.415 & 0.3\\
 4.00  & 1.4912  &$ $0.01& 11.382 & 0.0\\
 4.50  & 1.4914  &$ $0.03& 11.379 & 0.0\\
\end{tabular}
\caption{ Results for $r_{\mathrm{e}}$ and $f$ of the silane molecule.
Five basis spheres of the same radius $R$ centered on the atoms are
used. The cubic simulation cell has sides of length 12~\AA. $\lmax =
2$.  The experimental value of $r_{\mathrm{e}}$ is 1.4798~\AA, while
the equivalent plane-wave calculations give the values of 1.4910~\AA\
and 11.38~N/cm for $r_{\mathrm{e}}$ and $f$, respectively.  }
\label{tb_sih4}
\end{table}
%==================
\par Fig.~\ref{fig_sih4BL1_83} shows the total energy of the silane
molecule with a Si-H bond length of 1.83~\AA, as a function of
$E_{\mathrm{c}}$ and $R$.  Total energy differences converge for
cutoff energies above 800~eV.  The results of the calculations of
$r_{\mathrm{e}}$ and $f$ (for the breathing mode) are summarized in
Table~\ref{tb_sih4}.  We find that the accuracy is acceptable when $R
= $ 3.00~\AA.

%-------------
\begin{table}
\begin{tabular}{l|rrrr}
$R$(\AA)   & $r_{\mathrm{e}}$(\AA)  & $\delta r_{\mathrm{e}}(\%) $ &
$f$$\left(\frac{\mathrm{N}}{\mathrm{cm}}\right)$ & $\delta f(\%) $
\\
\hline
 2.00  & 1.4851 &$-$0.40& 11.847 & 4.1 \\
 2.50  & 1.4856 &$-$0.36& 11.948 & 5.0 \\
 3.00  & 1.4902 &$-$0.05& 11.438 & 0.5 \\
 3.50  & 1.4906 &$-$0.03& 11.419 & 0.3 \\
 4.00  & 1.4912 &$ $0.01& 11.382 & 0.0\\
\end{tabular}
\caption{ Results for $r_{\mathrm{e}}$ and $f$ of the silane molecule.  Five basis spheres
centered on the atoms are used.  The radius of the basis sphere
centered on the silicon atom is fixed at 4.00~\AA\ but the radius $R$
of basis spheres centered on the hydrogen atoms is varied.  }
\label{tb_sih4_varyRH}
\end{table}
%==================
\par We repeat the $r_{\mathrm{e}}$ and $f$ calculations on the silane
molecule with the radius of the basis sphere centered on the silicon
atom fixed at 4.00~\AA, but with the radius $R$ of the basis spheres
centered on the hydrogen atoms varied. The results in
Table~\ref{tb_sih4_varyRH} show that an accuracy of 1\% can be achieved
by using $R = 3.00$~\AA, which is 1~\AA\ larger than the basis spheres
centered on the hydrogen atom in the hydrogen chloride molecule
calculation (c.f. Table~\ref{tb_2_diff_sp_hcl_800eV_varyrH}).

%----------
\begin{figure}
%\centerline{\includegraphics[width=8cm,clip]{./fig.09.eps}}
\includegraphics[width=\FigureLength,clip]{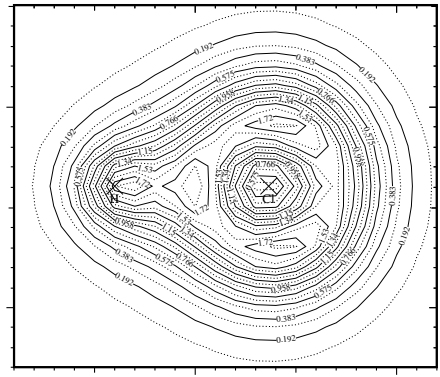}
\caption{ Electronic densities (in units of~electrons/\AA$^{3}$) on the plane
containing atoms of the hydrogen chloride molecule with a bond length
of 1.2746~\AA.  The locations of the atoms are marked by crosses.  }
\label{fig_hcl_density}
\end{figure}
%==================

%----------
\begin{figure}
%\centerline{\includegraphics[width=8cm,clip]{./fig.10.eps}}
\includegraphics[width=\FigureLength,clip]{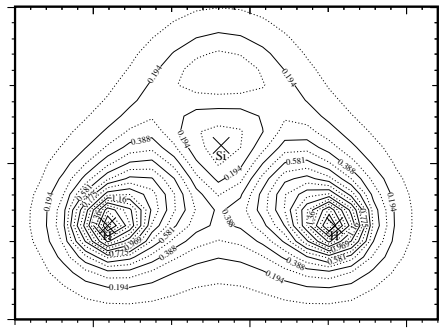}
\caption{ Electronic densities (in units of~electrons/\AA$^{3}$) on the plane
containing three atoms of the tetrahedral silane molecule with bond
lengths of 1.4798~\AA.  The locations of the atoms are marked by
crosses.  }
\label{fig_sih4_density}
\end{figure}
%==================
\par From the pseudo-charge density of the hydrogen chloride molecule
(Fig.~\ref{fig_hcl_density}), we observe that the valence electrons
are concentrated towards the chlorine atom, as expected from the
relative electronegativites of hydrogen and chlorine.  This enables us
to use a smaller basis sphere centered on the hydrogen atom to obtain
accurate results. However, from the pseudo-charge density of the
silane (Fig.~\ref{fig_sih4_density}), we observe substantial charge
density around the hydrogen atoms, reflecting the fact that hydrogen
is more electronegative than silicon.  Hence for the silane molecule
calculations, the radius of the basis spheres centered on the hydrogen
atoms need to be larger than that for hydrogen chloride.  These
observations lead to the conclusion that the relative
electronegativities of neighboring atoms in a calculation should be
taken into account when choosing basis sphere radii.

%----------
\begin{figure}
%\centerline{\includegraphics[width=8cm,clip]{./fig.11.eps}}
\includegraphics[width=\FigureLength,clip]{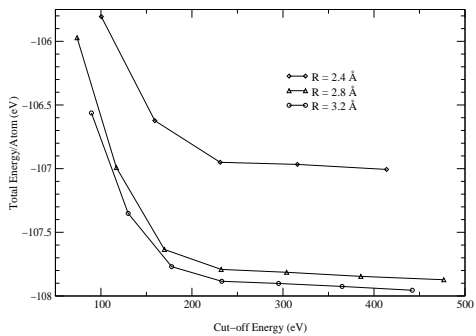}
\caption{ Total energy of a 64-atom cell of bulk crystalline silicon
with a lattice parameter of 5.43~\AA\ and $\lmax=2$.  64 basis spheres
of the same radius $R$ centered on the atoms are used.  }
\label{fig_si64_lc_5_43L2}
\end{figure}
%==================

\par We have chosen bulk crystalline silicon to test the performance
of the basis set on an extended system.  Fig.~\ref{fig_si64_lc_5_43L2}
shows the total energy per atom for a 64-atom cell of silicon with a
lattice parameter of 5.43~\AA\, as a function of the cutoff energy
and $R$.  The total energy converges at a cutoff energy of about
250~eV.  The rapid convergence of the total energy with respect to $R$
is evident from the figure.

%-------------
\begin{table}
\begin{tabular}{l|rrrr}
$R$(\AA)   & $a$(\AA)  & $\delta a(\%) $ &
$B$(GPa) & $\delta B(\%) $
\\
\hline
 2.60 & 5.353 &$-$0.78& 139.1& 50.7\\
 2.80 & 5.413 &$ $0.33& 104.5& 13.2\\
 3.00 & 5.445 &$ $0.93& 97.0 & 5.1\\
 3.20 & 5.453 &$ $1.08& 89.8 & $-$2.7\\
\end{tabular}
\caption{ Results for the equilibrium lattice parameter $a$ and bulk
modulus $B$ of the 64-atom bulk crystalline silicon, with $\lmax=1$.
64 basis spheres of the same radius $R$ centered on the atoms are
used. The experimental values for $a$ and $B$ are 5.43~\AA\ and
100.0~GPa, respectively. The plane-wave calculations, with a cutoff
energy of 250~eV, give the results of 5.395~\AA\ and 92.3~GPa,
respectively.  }
\label{tb_silmax1}
\end{table}
%==================

%-------------
\begin{table}
\begin{tabular}{l|rrrr}
$R$(\AA)   & $a$(\AA)  & $\delta a(\%) $ &
$B$(GPa) & $\delta B(\%) $
\\
\hline
 2.60 & 5.310&$-$1.58&238.0& 157.9\\
 2.80 & 5.353&$-$0.78&129.8& 40.6 \\
 3.00 & 5.377&$-$0.33&111.4& 20.7\\
 3.20 & 5.385&$-$0.19&96.2 & 4.2\\
\end{tabular}
\caption{ Results for the equilibrium lattice parameter $a$ and bulk
modulus $B$ of the 64-atom bulk crystalline silicon, with $\lmax=2$.
64 basis spheres of the same radius $R$ centered on the atoms are
used.  }
\label{tb_silmax2}
\end{table}
%==================
\par To determine the equilibrium lattice parameter, $a$, and the bulk
modulus, $B$, we perform a series of calculations on the bulk silicon
system for a range of lattice parameters from 5.31~\AA\ to 5.51~\AA.
The results of the calculations with $\lmax=$ 1 and 2 are tabulated in
Tables~\ref{tb_silmax1} and \ref{tb_silmax2}, respectively.  It is
found that even with $\lmax=1$, the results with $R=3.20$~\AA\ agree quite
well with the plane-wave and experiment results.  The calculations
with $\lmax=2$ improve the results slightly.  The reason why $\lmax=1$
calculations give rather good results is because silicon atoms mix the
$s$ and $p$ states to form four $sp^3$ orbitals which are obviously
well-described by a basis set with $\lmax=1$.

%-------------
\begin{table}[h]
\begin{tabular}{l|rrr}
System & $E_{\mathrm{c}}$(eV) &  $N_{\scriptscriptstyle\mathrm{SW}} $ &  $N_{\scriptscriptstyle\mathrm{PW}}$
\\
\hline
 Chlorine molecule  & 800     & 234  & 88663\\
 Hydrogen molecule  & 1000    & 270  & 124097\\
 HCl molecule       & 800     & 288  & 88663\\
 Silane molecule    & 800     & 720  & 88663\\
 Bulk silicon       & 250     & 4608 & 10827\\
\end{tabular}
\caption{Numbers of the basis functions required so
that the agreement between the results for $r_{\mathrm{e}}$ and $f$ (for
molecules); and $a$ and $B$ (for bulk silicon) from the spherical-wave
and plane-wave basis set calculations is less than 1\%.
The numbers of the spherical-wave and plane-wave basis functions 
are denoted by $N_{\scriptscriptstyle\mathrm{SW}}$ and 
$N_{\scriptscriptstyle\mathrm{PW}}$, respectively. For the
spherical-wave basis set calculations, the choice of atom-centered 
and $\lmax = 2 $ is used.
$E_{\mathrm{c}}$ is the cutoff energy.}
\label{tb_basisfunctionnumber}
\end{table}
%==================

Finally we present Table~\ref{tb_basisfunctionnumber} which shows the
numbers of basis functions for the spherical-wave
and plane-wave basis set calculations. 
Since in general the number of spherical-wave basis functions is very small 
for molecules compared to that of plane-wave basis functions, 
we conclude that
spherical-wave basis sets can be used to study molecules and possibly
clusters with high efficiency.

\section{Conclusions}
\label{sec_summary}
\par Through detailed calculations on molecules and bulk crystalline
silicon, we find that the total energy and physical properties can be
accurately deduced from density-functional calculations using
localized spherical-wave basis sets.  We find that for most purposes,
the choice of $\lmax=2$ and atom-centered basis spheres is sufficient
to obtain an accuracy which is excellent compared to those obtained
using the extended plane-wave basis set.

\par The dependence of the total energy on the cutoff energy of the
localized basis set is found to be rather similar to that of the
extended plane-wave basis set.  We also find that the results converge
exponentially with respect to the basis sphere radius $R$.
The angular incompleteness can be improved either by increasing
$\lmax$, or by introducing additional basis spheres at strategic
locations (such as at the middle of a bond).

\par We find that the radii of the basis spheres depend on the bond
lengths.  A large bond length usually means large basis spheres are to
be used.  We find that it is possible to use different radii for the
basis spheres depending on the relative electronegativities of the
atomic species.  Usually we need to use basis spheres with larger
radii for atoms that are more electronegative than other atoms in a
calculation.

\par For the bulk silicon case, the accuracy of the results obtained
using $\lmax=1$ is marginally acceptable, which is a consequence of
the $sp^3$ hybridization.

\par Finally, we note that one of the main advantages of this basis
set is that the accuracy of the results can be systematically
improved. It would be interesting to explore the possibility of 
using two or more basis spheres of different radii
centered on an atom so that a lower cutoff energy for the larger basis
spheres might be used.

\section*{Acknowledgement}
\par C.K.G.\ acknowledges financial support from the Cambridge Commonwealth
Trust and from St.~John's College, Cambridge. P.D.H.\ acknowledges the
support of a Research Fellowship from Magdalene College, Cambridge.
\bibliographystyle{prsty}

\end{document}